\documentclass[fleqn,usenatbib]{mnras}
\usepackage[usenames,dvipsnames]{xcolor}

\usepackage[T1]{fontenc}
\usepackage{ae,aecompl}

\usepackage{graphicx}% Including figure files
\usepackage{amsmath}	% Advanced maths commands
\usepackage{amssymb}	% Extra maths symbols
\graphicspath{{./images/}}

\title[Constraints on WD 1145+017]{Mass and eccentricity constraints on the planetary debris orbiting the white dwarf WD 1145+017}

\author[Gurri,Veras and G\"{a}nsicke]{
Pol Gurri,$^{1,2}$\thanks{E-mail: polgurri@gmail.com}
Dimitri Veras,$^{2}$
and Boris T. G\"{a}nsicke$^{2}$
\\
$^{1}$ Departament de F\'{i}sica, Universitat Polit\`{e}cnica de Catalunya, c/Esteve Terrades 5, 08860 Castelldefels, Spain\\
$^{2}$Department of Physics, University of Warwick, Coventry CV4 7AL, UK\\
}

\date{Accepted 2016 September 08. Received 2016 September 06; in original form 2016 June 21.}

\pubyear{2016}

\begin{document}
\label{firstpage}
\pagerange{\pageref{firstpage}--\pageref{lastpage}}
\maketitle

\begin{abstract}
Being the first of its kind, the white dwarf WD\,1145+017 exhibits a complex system of disintegrating debris which offers a unique opportunity to study its disruption process in real time. Even with plenty of transit observations there are no clear constraints on the masses or eccentricities of such debris. Using $N$-body simulations we show that masses greater than $\simeq10^{20}$\,kg (a tenth of the mass of Ceres) or orbits that are not nearly circular ($\mathrm{eccentricity}>10^{-3}$) dramatically increase the chances of the system becoming unstable within two years, which would contrast with the observational data over this timespan. We also provide a direct comparison between transit phase shifts detected in the observations and by our numerical simulations.

\end{abstract}


%%%%%%%%%%%%%%%%% BODY OF PAPER %%%%%%%%%%%%%%%%%%

\section{Introduction}

Planets which survive the giant branch evolution of their hosts stars are expected to be rather common \citep{burleigh2002,villaver2007,mustil12,veras13}. This prediction is corroborated by the detection of photospheric metal pollution in a large fraction of all  white dwarfs  \citep{zuckerman03, zuckerman2010, koester14}. Dynamical interactions in evolved planetary systems can scatter planetary bodies near the Roche radii of the white dwarfs \citep{debes02,frewen2014,payne2016a,payne2016b} where they are tidally disrupted \citep{jura2003,debes2012,veras14a,veras15c}, forming detectable accretion discs \citep{zuckerman1987,boris2006,kilic2006,farihi2009,bergfors2014}, and ultimately accreting onto the white dwarf.  Analysis of the photospheric trace metals provides detailed insight into the bulk chemical compositions of planetary systems \citep{zuckerman2007,boris2012,xu2014}, which in turn guides planet formation models (e.g. \citealt*{bond2012}). The current observational and theoretical progress on evolved planetary systems is summarised by \citet{farihi2016} and \citet{veras2016rev}.

\cite{vanderburg1} announced transits recurring with a period of $\simeq4.5$\,h in the $K2$ light curve of the white dwarf WD\,1145+017, which also exhibits infrared excess from a circumstellar disk and photospheric metal pollution. The orbit of the transiting objects lies close to the disruption, or Roche, limit for rocky objects. Thus, WD\,1145+017 represents the first observational detection of planetesimals orbiting a white dwarf, opening a new window into the understanding of poorly-known processes such as disintegration, orbital circularisation, or the actual nature of those orbiting bodies \citep{veras14b,veras15b,veras2016rev}.

In this work we derive constraints for the masses and eccentricities of the bodies orbiting the star from N-body simulations of the system. Section \ref{sec:system}, provides an overview of the WD\,1145+017 system and Sect. \ref{sec:simulation} outlines the setup of our simulations. In Sect. \ref{sec:mass} and \ref{sec:ecc} we present the results that set constraints on the mass and eccentricity of the orbiting debris. Section \ref{sec:shifts} is devoted to phase shifts, proving also a direct comparison between observational data and our simulations.  We briefly discuss our results in Sect. \ref{sec:disc} and then conclude in Sect. \ref{sec:conc}.

\section{The WD\,1145+017 system} \label{sec:system}

The 17th magnitude white dwarf WD\,1145+017, first identified by \cite{berg1992} and rediscovered by \cite{friedrich2000}, was observed during Campaign\,1 of the extended \textit{Kepler} mission, and \citet{vanderburg1} discovered transits of at least one, and probably several, bodies with periods ranging from 4.5\,h to 4.9\,h in the \textit{K2} light curve. Deep ($\simeq40$\%) transits lasting $\sim10$\,min recurring every 4.5\,h (near the Roche-limit for a rocky body) were confirmed in ground-based follow-up photometry, which \citet{vanderburg1} interpreted as dust and gas emanating from a smaller, undetected object, analogous to a cometary tail \citep{vanderburg1}. Optical spectroscopy revealed both photospheric metal pollution \citep{vanderburg1} and absorption from circumstellar gas \citep{xu16}, and an infrared excess detected in the UKIDSS and \textit{WISE} photometry confirmed the presence of circumstellar dust \citep{vanderburg1}.

High-speed photometry obtained by \citet{boris16} over the course of 15 nights in November and December 2016 revealed the rapid evolution of the system since the discovery by \cite{vanderburg1}. Multiple transits typically lasting $3-12$\,min and with depths of $10-60$\% were observed on every occasion. While these transit events changed depth and shape on time scales of days, \citet{boris16} could track six individual features over at least three individual observing nights, and concluded that at least six objects were orbiting WD\,1145+017 on nearly identical orbits, with a mean period of $4.4930\pm0.0013$\,h. This period is significantly distinct, and shorter, compared to the dominant 4.5\,h period measured by \citet{vanderburg1} from the \textit{K2} data. A few transits with similarly short periods were also detected by \citet{croll15}.

Additional extensive photometry obtained with small-aperture telescopes confirmed the presence of multiple objects with periods of $\simeq4.493$\,h, as well as the 4.5\,h period detected in the \textit{K2} data \citep{rappaport16}. Comparing these two distinct periods, \citet{rappaport16} used an analytical model in which fragments drift off a Roche-lobe filling asteroid to estimate the mass of the asteroid to be $\simeq10^{20}$\,kg, about one tenth of the mass of Ceres. In the context of this paper, we will refer to the object at $\simeq4.5$\,h as the \textit{parent body}, and to the multiple objects with periods of $\simeq4.493$\,h as \textit{fragments}.

The physical nature of the obscuring material has been investigated by \citet{alonso16} who obtained spectroscopic transit observations through a wide slit. Binning their data into four bands centred at 0.53, 0.62, 0.71, and 0.84\,$\mu$m, \citet{alonso16} found practically no colour-dependence of the transit shapes and depths, and concluded that the particle size of the debris must be $\ga0.5\mu$m. More recently, \citet{zhou16} obtained multi-band photometry spanning $0.5-1.2\mu$m using several telescopes, and derived a $2\sigma$ lower limit on the particle size of $0.8\mu$m.

Several estimates of the current accretion rates were derived from the dust extinction ($\simeq8\times10^6\,\mathrm{kg\,s^{-1}}$, \citealt{vanderburg1}; $\simeq10^8\,\mathrm{kg\,s^{-1}}$, \citealt{boris16}) and the gas absorption lines ($\simeq10^9\,\mathrm{kg\,s^{-1}}$, \citealt{xu16}). The metal content of the white dwarf envelope is $\simeq6.6\times10^{20}$\,kg \citep{xu16}, however, given that the time scales on which the metals diffuse out of the envelope are a few $10^5$\,yr, it is not possible to unambiguously associate that metal content with the ongoing disruption event.

\section{Simulation Setup} \label{sec:simulation}

Our $N$-body numerical simulations are based on a model of WD\,1145+017 that, while taking into account the results from the recent follow-up observations, ignores the detailed process of the tidal disruption of the bodies orbiting the white dwarf \citep{debes12,veras14a}, as well as effects such as collisions and interactions with the debris disk, white dwarf radiation \citep{veras14b,veras15a,veras15b} or gas drag \citep{veras15a,veras15c} which may play an important role in the evolution of the system. The study of \cite{veras2016a} considered only equal-mass bodies in initially strictly co-orbital configurations, an approach which is unsuitable for WD\,1145+017. We rather intend to provide an understanding of many-body interactions without these restrictions, and specifically for the system harbouring WD\,1145+017. 

All simulations are performed using the $N$-Body code {\sc Mercury} \citep{chambers99} with some extra modifications which include the effects of general relativity and improve the collision detection mechanism in the same way as used by \cite{veras13}. Our simulations record the interactions between bodies for a duration of at least two years, spanning approximately the observational baseline. We use a BS (Bulirsch-Stoer) integrator with an accuracy parameter of $10^{-12}$. For every set of simulations, a subset of them are re-run with a smaller accuracy parameter of $10^{-13}$ and results are double-checked using both Cartesian and Keplerian coordinates.

Each simulation recorded the interactions between a parent body at an orbital period of 4.5004h and six fragments at 4.493\,h orbiting a $0.6$M$_{\odot}$ and $1.4$R$_{\oplus}$ white dwarf like WD\,1145+017 \citep{vanderburg1}. 
We chose to simulate 6 fragments because: (i) simulating only one fragment would neglect fragment-fragment interactions that likely play an important role (e.g. \cite{veras2016a}), (ii) observations reveal that there are at least 6 fragments \cite{boris16,rappaport16} and (iii) adopting many more fragments would have been computationally challenging. We placed the fragments on the same shorter-period orbit relative to the parent body in order to (i) compare our numerical results with the analytical model of \cite{rappaport16}, (ii) avoid an unnecessary and expensive exploration of phase space, (iii) match the results of \cite{veretal2016b}, which demonstrate through disruption simulations that multiple fragments settle into shorter-period orbits.

The lack of strong observational evidence supporting any specific orbital configuration for the bodies motivated us to randomly sample from a uniform distribution the initial mean anomalies for all bodies and simulations. Also, under the hypothesis of the fragments being tidally-disrupted parts of the parent body, we assumed their masses to range from 0.01\% to 20\% of the parent body's mass, and we lineally randomised the masses of all fragments for every simulation between those values. Therefore, the only free, non-random parameter in our simulations was the mass of the parent body.

In our simulations we have found that there are two different types of interactions that can vary the orbit of the fragments: (i) perturbations caused by the parent body and (ii) perturbations produced by other fragments. Fragment-fragment interactions are mainly responsible for small orbital dispersions (i.e. period deviations < 10s) coinciding with the findings of \cite{veretal2016b}. These smaller period deviations can reduce the radial distance between fragments and parent, and therefore strengthen fragment-parent interactions which contribute greatly to period deviation. Fragment-parent interactions are of special interest when the fragment and the parent are closer together ($\simeq$ every 100 days) and they result in short-term changes in the eccentricity of the fragment. Figure \ref{fig:3trajectories} displays repetitive peaks in the fragment's period deviation due the short-term eccentricity deviations. Fragment-parent interactions depend greatly on the eccentricity of the parent body, to the point that great parent eccentricities ($>10^{-2}$) result in most of the fragments colliding with the parent body. 

\section{Mass constraints} \label{sec:mass}

Being able to obtain an order-of-magnitude constraint on the mass of the bodies was the first motivation of the simulations. Observations suggest that the planetesimal(s) at WD\,1145+017 have been in the same, or very similar, orbits for at least $\sim 500$ days \citep{boris16} and therefore should have low eccentricities (tidal disruption strongly correlates with orbital pericentre instead of semi-major axis \citep{veras14b}, meaning that highly eccentric orbits would cause a rapid disruption of planetesimals). Thus, we imposed completely circular orbits for all bodies and ran simulations for an extended time of five years to gain insight into the time scale on which the system evolves. We gradually increased the parent body's mass, using 10 linearly-distributed values per decade in mass, from $\sim 10^{18}$\,kg to $\sim 10^{21}$\,kg and performed 100 simulations for each mass value. As a result, we obtained detailed trajectories for all bodies over time, allowing us to track changes in their orbital periods.

\begin{figure}
	\includegraphics[width=\columnwidth]{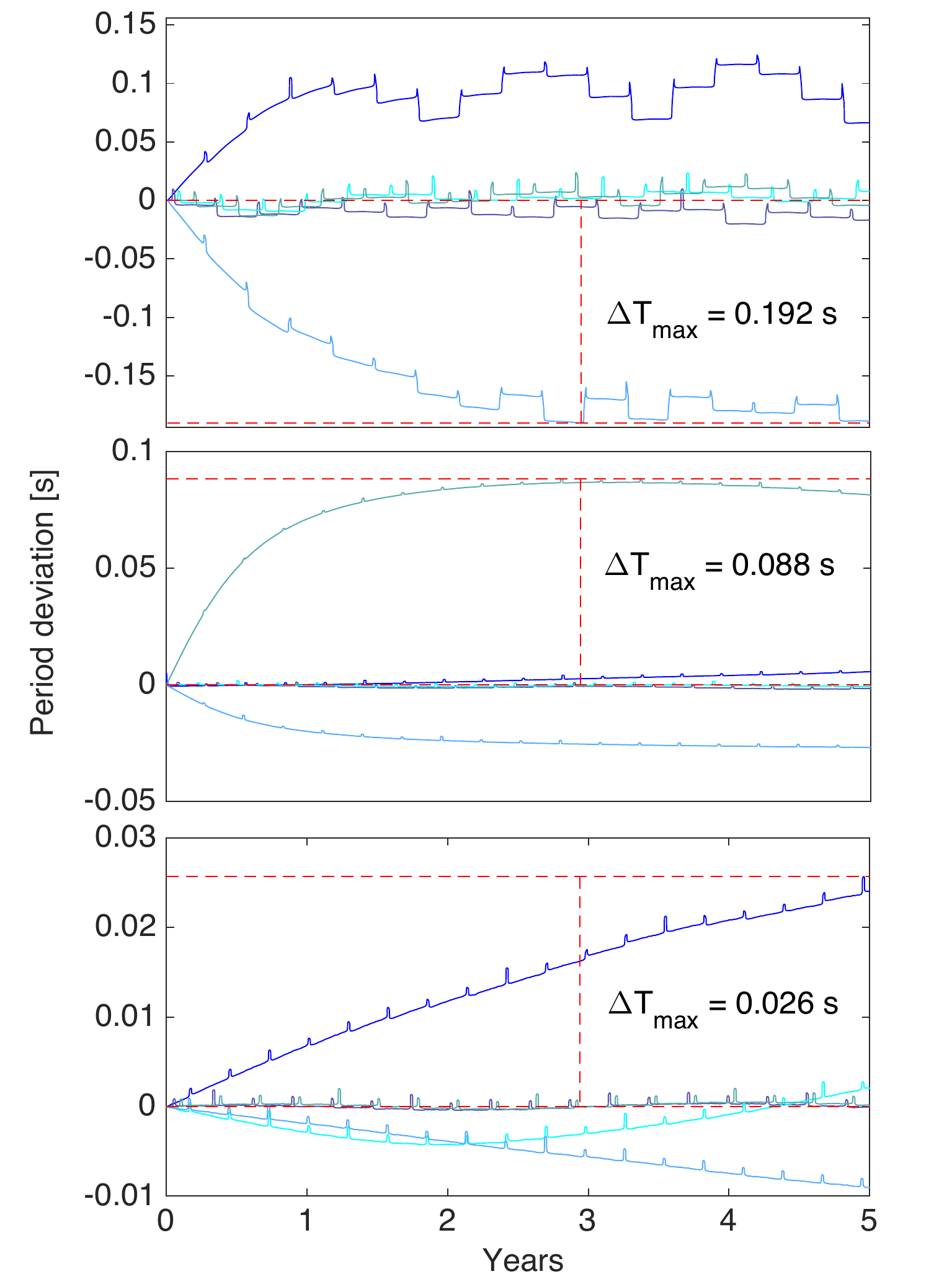}
    \caption{Each line represents the period deviation $\Delta T(t) = T(t)-T(0)$ of a fragment as a function of time. Three different simulations are shown to help visualize the complex behaviour of fragments' trajectories and to illustrate our definition of the period deviation $\Delta T_{\rm max}$. The mean $\langle \Delta T_{\rm max}\rangle$ is computed as the average of all values of $\Delta T_{\rm max}$, while Max$(\Delta T_{\rm max})$ and Min$(\Delta T_{\rm max})$ are the highest and lowest values among all $\Delta T_{\rm max}$.}
    \label{fig:3trajectories}
\end{figure}

\begin{figure}
	\includegraphics[width=\columnwidth]{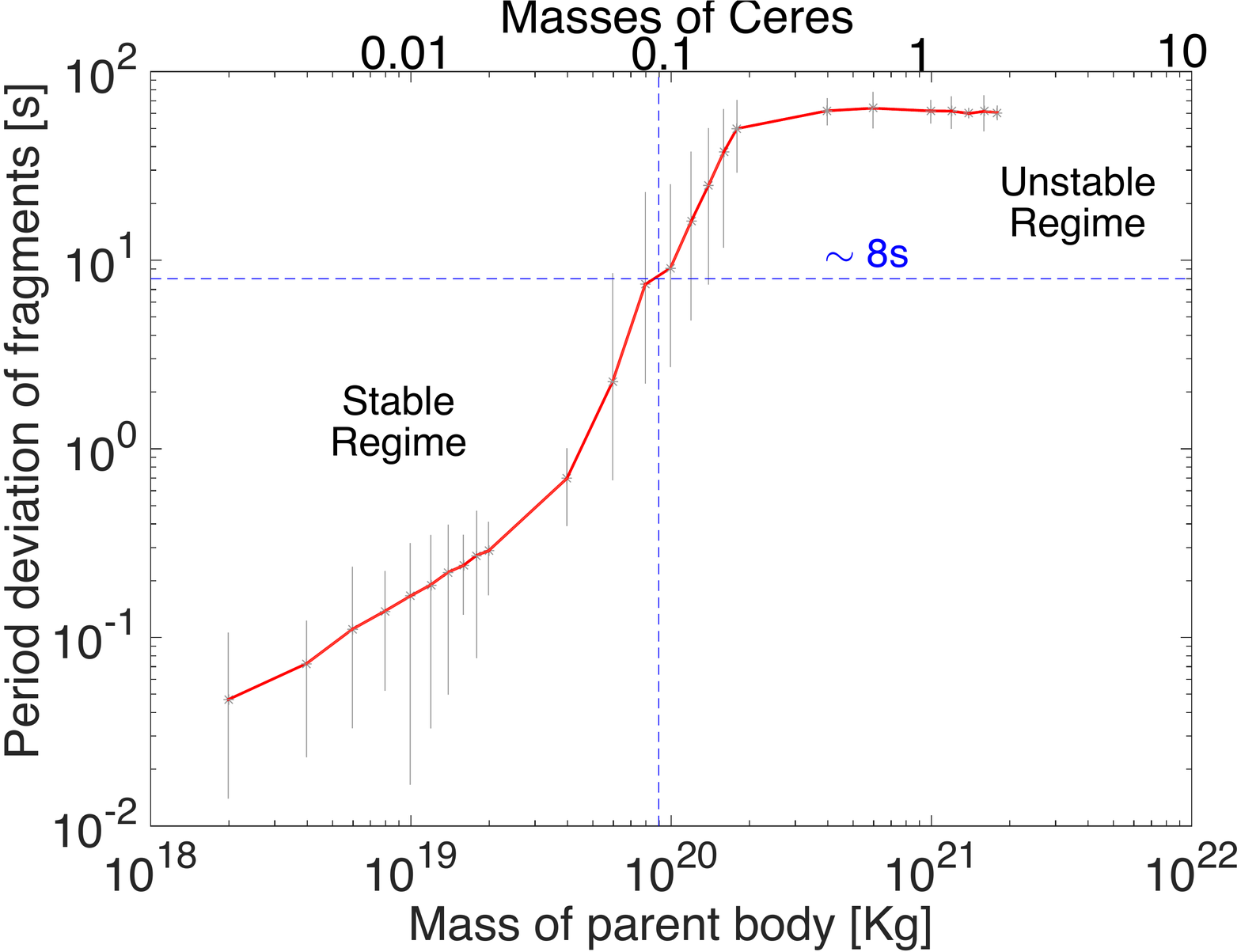}
    \caption{Mean of maximum fragment orbital period deviation, $\langle \Delta T_{\rm max}\rangle$ (see Section \ref{sec:mass}), for different parent body masses. Each point represents the mean of 100 simulations under the assumptions of completely circular orbits and fragment masses randomly ranging from 0.01\% to 20\% of the parent body's mass. Each simulation spans 5 years ($\simeq9750$ fragment's orbital cycles) .Blue dashed lines are the theoretical prediction for the mass of the parent body \citep{rappaport16} and the maximum difference between the observed orbital periods and their mean \citep{boris16}. Only low-mass systems tent to exhibit small interactions and orbital stability over time.}
    \label{fig:circularorbperiod}
\end{figure}

Interactions between bodies can cause observationally measurable deviations in the orbital periods \citep{veras2016a}. Therefore, we monitored the difference that each body experienced between its orbital period during the simulation, $T(t)$, and its initial orbital period, $T(0) = 4.493$\,h, which we will refer as $\Delta T(t) = T(t)-T(0)$. For each simulation, we only kept the maximum value of $\Delta T(t)$ among six fragments, $\Delta T_{\rm max}$. Figure \ref{fig:3trajectories} displays the period deviation of fragments for 3 different simulations and clarifies the meaning of $\Delta T_{\rm max}$. Performing 100 simulations for each value of parent body's mass yielded an average maximum period deviation $\langle \Delta T_{\rm max}\rangle$ which depended only on the mass of the parent body.

$\Delta T_{\rm max}$ provides insight into the stability of the system. Large values of $\Delta T_{\rm max}\gtrsim 20$\,s translates into fragments being greatly affected by the parent body and experiencing potentially detectable changes in their orbital periods. With the periods of the fragments and the parent body being initially separated by $\simeq 30$\,s, period deviations up to 60\,s resulted in some simulations where fragments entered new orbits around the parent body! In contrast, small period deviations imply a low level of interaction between bodies and a greater chance of long-term dynamical stability. 

Figure \ref{fig:circularorbperiod} plots $\langle \Delta T_{\rm max}\rangle$ with respect to the mass of the parent body. Grey lines give the dispersion of the obtained $\Delta T_{\rm max}$ measures. The vertical blue dashed line is the \cite{rappaport16} theoretical prediction for the mass of the parent body and the horizontal blue dashed line is representing the maximum difference between the observed orbital periods and their mean \citep{boris16}.

Although the maximum difference between the observed orbital periods and their observational mean (horizontal blue dashed line in Figure \ref{fig:circularorbperiod}) does not directly relate with period deviation, it provides a sense of how radially distant are fragment's orbits in stable configurations. Because no fragment has been observed further than $\simeq 8$\,s away from the mean orbital period, it can set an order-of-magnitude upper limit for $\langle \Delta T_{\rm max}\rangle$. One caveat is that a large $\Delta T_{\rm max}$ can be generated by one fragment alone, which might be undetectable observationally. We should also note that although observations suggest a nearly constant orbital period of the fragments over $\simeq 500$\,days, $\Delta T_{\rm max}$ in our simulations is computed over five years. We aimed to reduce these uncertainties by averaging over 100 simulations.

Figure \ref{fig:circularorbperiod} shows a clear positive correlation between $\langle \Delta T_{\rm max}\rangle$ and the parent body's mass, with the strongest dependence occurring between $5\times 10^{19}$\,kg and $2\times 10^{20}$\,kg and then levelling off at about $5\times 10^{20}$\,kg. Assuming that $\langle \Delta T_{\rm max}\rangle$ has to be less than the difference between the observed orbital periods and their mean, this sharp increase provides a robust upper limit the parent body's mass, which is close to one tenth of the mass of Ceres, and agrees well with the estimate of the parent body mass analytically derived by \cite{rappaport16}.

\section{Eccentricity constraints} \label{sec:ecc}

As well as mass, eccentricity is very likely to play an important role in the stability of the system. We therefore ran simulations modifying not only mass but also the eccentricity of the parent body.

By keeping track of $\Delta T_{\rm max}$ in the same way as outlined in Section \ref{sec:mass}, we increased the parent body's mass from $\sim 10^{18}$\,kg to $\sim 10^{21}$\,kg and its eccentricity from $10^{-3}$ to $10^{-1}$. To obtain a $\langle \Delta T_{\rm max}\rangle$ we averaged over 50 two-year simulations per each pair of mass and eccentricity, henceforth denoted as a $(M,e)$ pair.

\begin{figure}
	\includegraphics[width=\columnwidth]{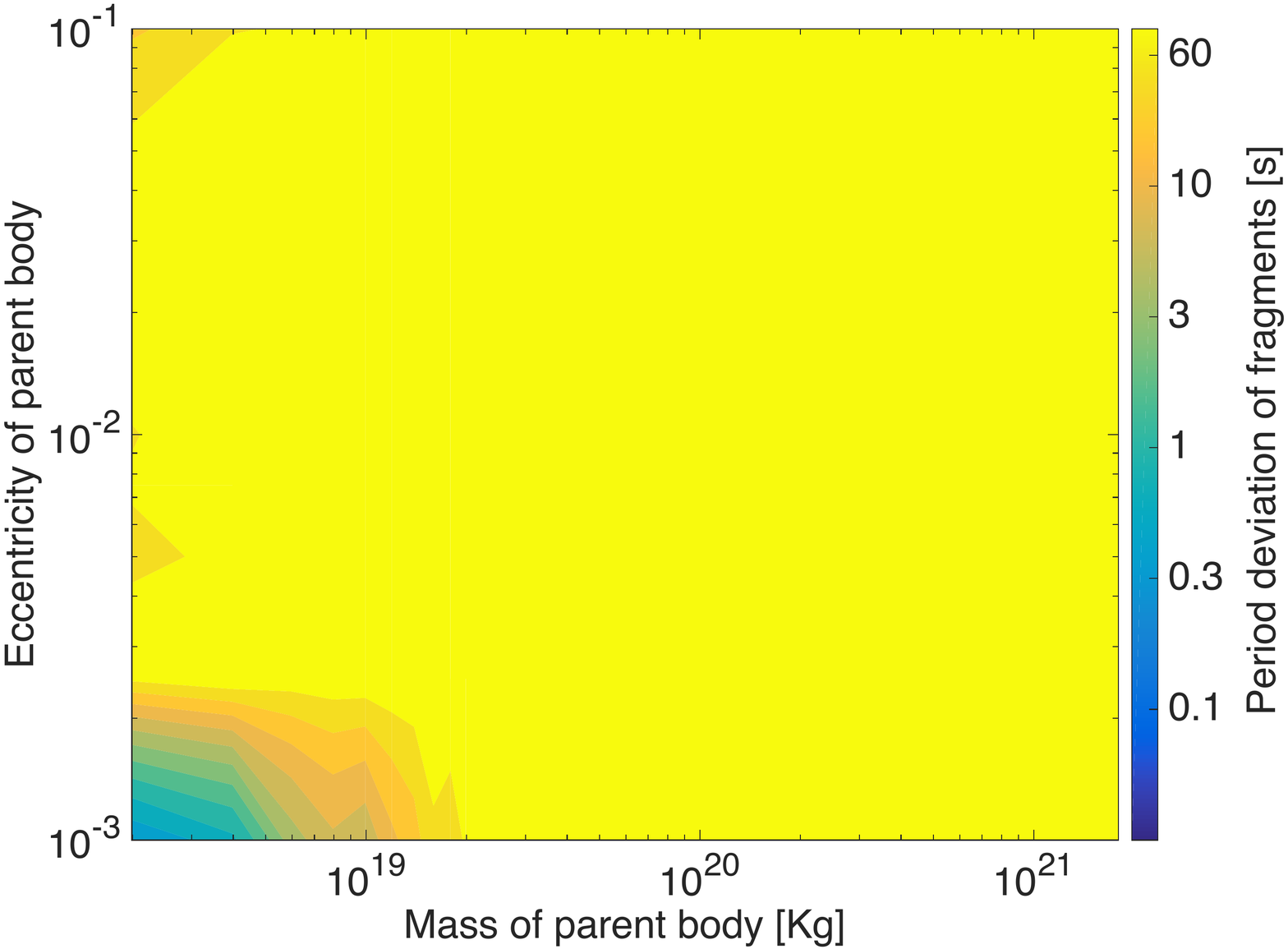}
    \caption{Mean of maximum fragment orbital period deviation, $\langle \Delta T_{\rm max}\rangle$ (see Section \ref{sec:mass}), as a function of the parent body's mass and eccentricity. Eccentricity and mass are sampled by 20 and 15 values, respectively, which are logarithmically spaced, and the fragment masses are randomly drawn ranging from 0.01\% to 20\% of the parent body's mass.  Each $(M,e)$ pair represents the mean of 50, 2-year simulations. Low values of $\langle \Delta T_{\rm max}\rangle$ are only found for both low parent body masses and low eccentricity.} 
    \label{fig:deviationfirstexploration}
\end{figure}

\begin{figure}
	\includegraphics[width=\columnwidth]{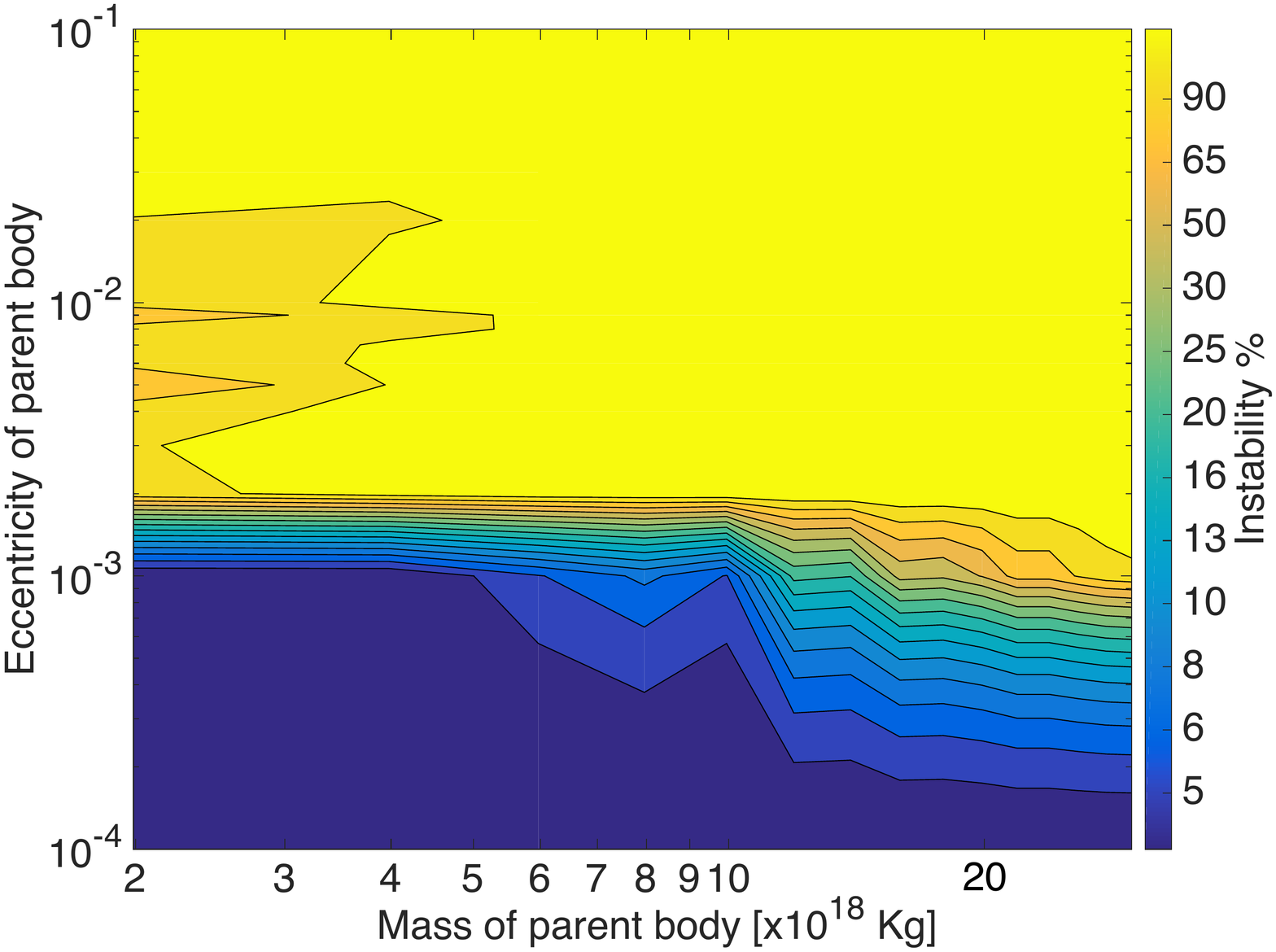}
    \caption{The fraction of unstable systems is defined as the ratio of 50 stable simulations over the total number of simulations performed (see Section \ref{sec:ecc}), and it is shown as function of the parent body's mass and eccentricity. Stable systems are defined as systems without collisions between fragments and parent or the star. There is a clear increase in the number of unstable systems with larger eccentricities.}
    \label{fig:instabilityHDD}
\end{figure}

\begin{figure}
	\centering
	\begin{minipage}[b]{\columnwidth}
  		\centering
  		\includegraphics[width=\columnwidth]{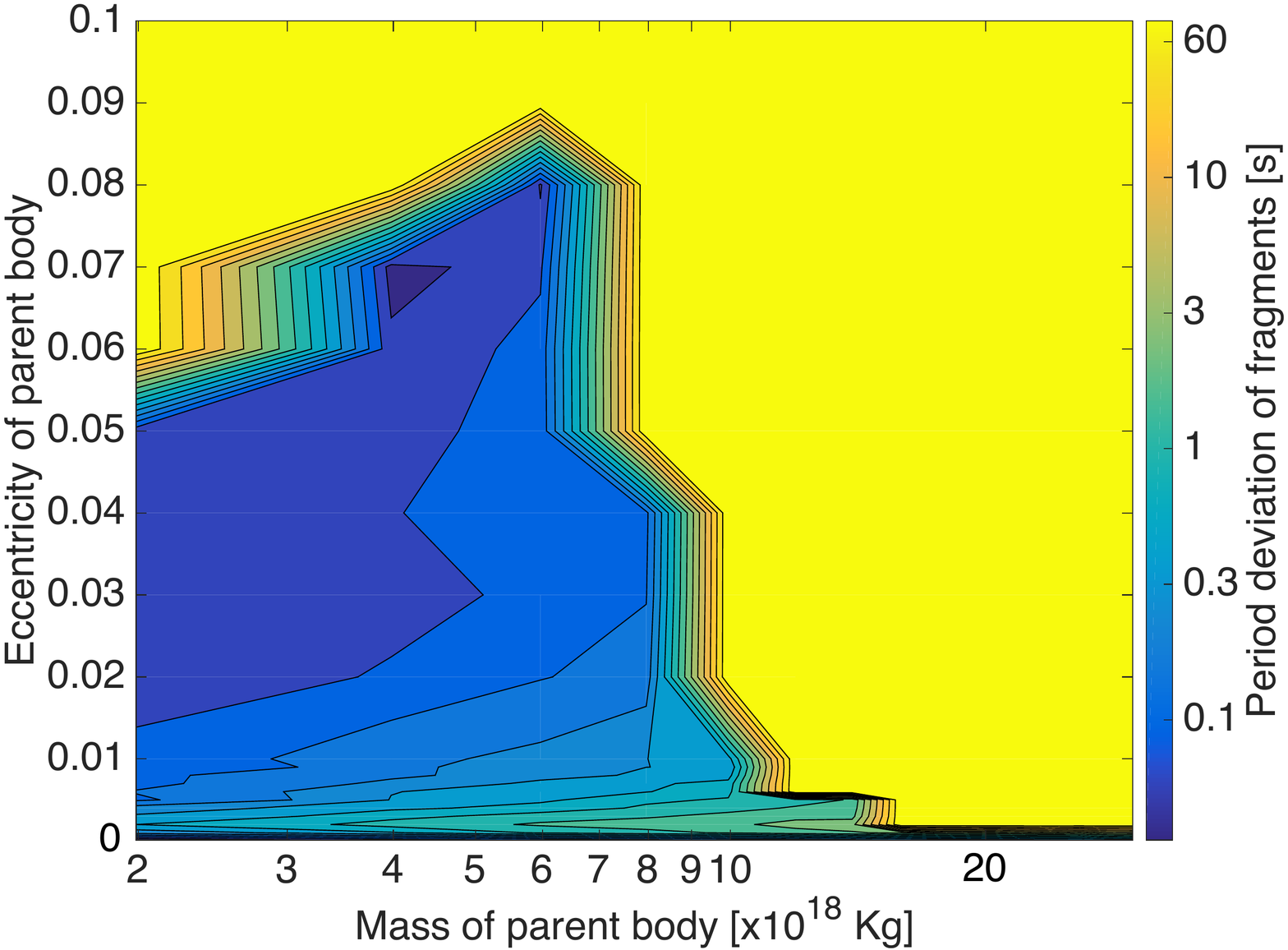}
	\end{minipage}
		
	\begin{minipage}[b]{\columnwidth}
  		\centering
  		\includegraphics[width=\columnwidth]{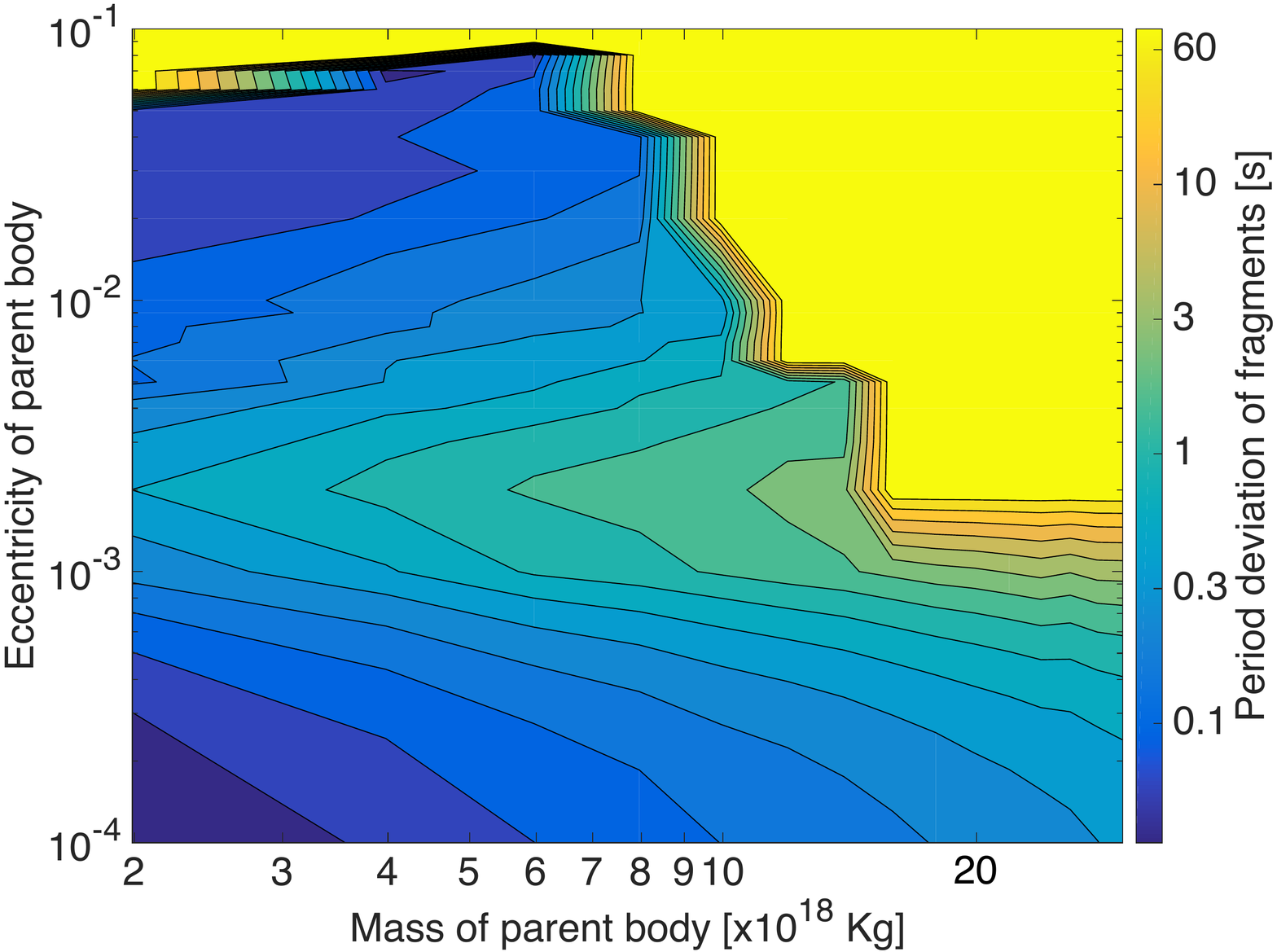}
	\end{minipage}
    \caption{Mean of maximum fragment orbital period deviation, $\langle \Delta T_{\rm max}\rangle$ (see Section \ref{sec:mass}), averaged for 50 stable simulations per $(M,e)$ pair (see Section \ref{sec:ecc}), shown on a linear scale in the top panel and on a logarithmic scale below. While systems with high eccentricities of up to $\simeq0.1$ are rarely stable (Fig.\,\ref{fig:instabilityHDD}), they can have low  $\langle \Delta T_{\rm max}\rangle$.}
    \label{fig:periodlog}
\end{figure}

\begin{figure*}
	\includegraphics[width=2\columnwidth]{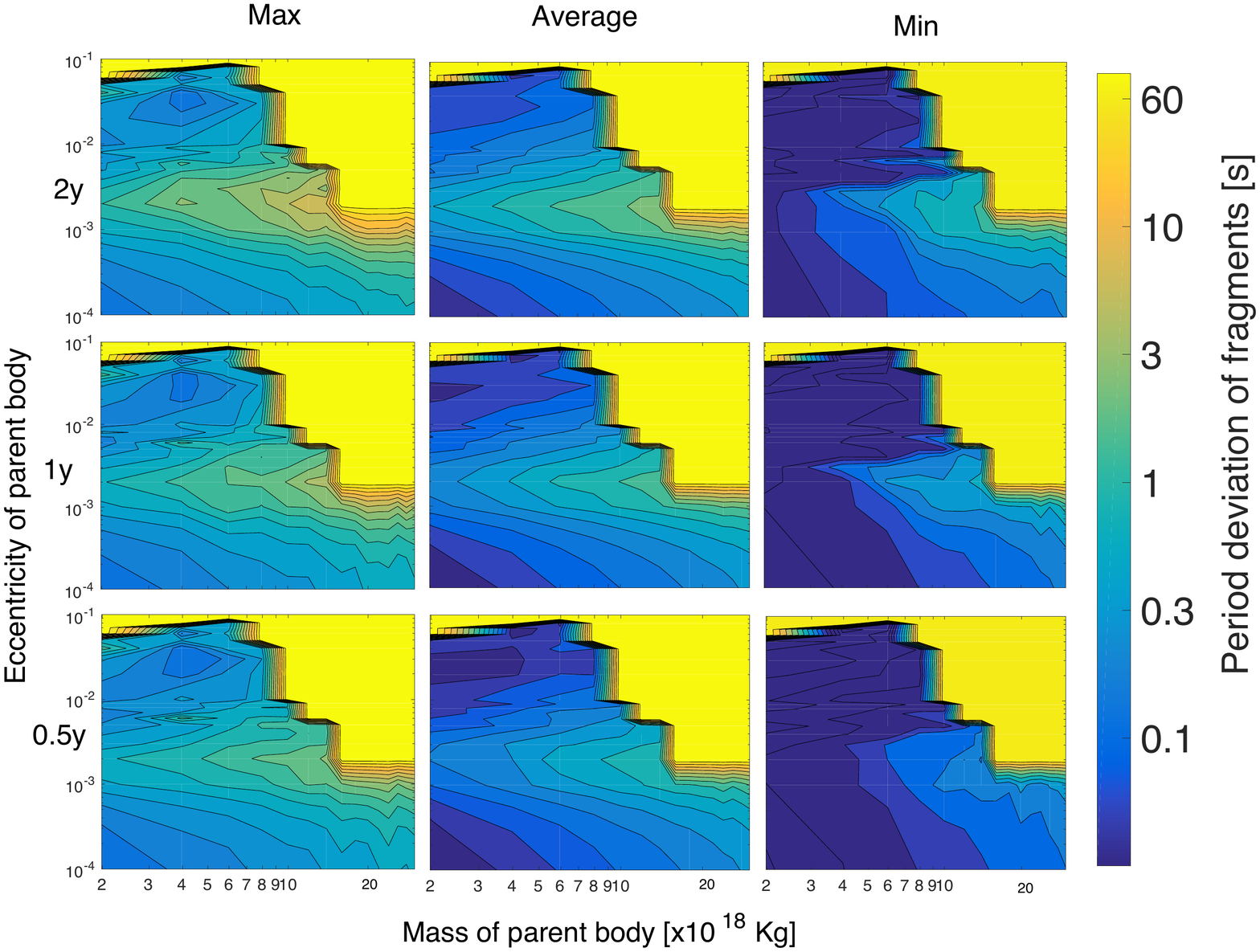}
    \caption{Max$(\Delta T_{\rm max})$, $\langle \Delta T_{\rm max}\rangle$ and Min$(\Delta T_{\rm max})$ for 50 stable simulations (see Section \ref{sec:ecc}), as a function of the parent body's mass and eccentricity. All plots are repeated for three different epochs throughout the simulations: six months, one year and two years after the start of the simulation. The number of analysed values as well as assumptions are the same than in Figure \ref{fig:deviationfirstexploration}. This figure illustrates the long-term robustness of stable systems over extended periods of time.}   
    \label{fig:maxaveminperiods}
\end{figure*}

Figure \ref{fig:deviationfirstexploration} exhibits contours which display $\langle \Delta T_{\rm max}\rangle$ as a function of mass and eccentricity of the parent body. A clear dependence on eccentricity can be seen, demonstrating that, (i) as well as mass, eccentricity plays a key role in period deviation, and (ii) that only systems with low mass and eccentricity exhibit small period deviations.

Because low period deviations agree with observations, the area of greatest interest of Figure \ref{fig:deviationfirstexploration} is the lower-left corner. Consequently, we performed extra simulations to analyse in more detail the region from $2\times 10^{18}$\,kg to $2\times 10^{19}$\,kg and sampled eccentricities in the range $10^{-4}$ to $10^{-1}$. We introduced the concept of ``stable systems'' to be able to study in detail those configurations that remain largely unperturbed over the entire time span of the simulation. We define a system as stable if none of the fragments collides with either the parent body or the star. With this definition, we run simulations until exactly 50 remain stable per $(M,e)$ pair. We kept track of the number of unstable simulations to be able to define the fraction of unstable systems per $(M,e)$ pair, defined as the ratio of unstable simulations ($N_\mathrm{unstable}$) divided by the total number of simulations performed ($N_\mathrm{unstable}+50$).

Figure \ref{fig:instabilityHDD} plots the fraction of unstable systems as a contour plot. A clear positive correlation can be seen between fraction of unstable systems and eccentricity, with a sharp increase in instability for eccentricities greater than $10^{-3}$. Mass has a less drastic effect but also increases instability, more evidently for masses greater than $10^{19}$\,kg.

Figure \ref{fig:periodlog} plots $\langle \Delta T_{\rm max}\rangle$ of the 50 stable solutions found per $(M,e)$ pair, both on linear and logarithmic eccentricity scales. A comparison between Figures \ref{fig:instabilityHDD} and \ref{fig:periodlog} highlights how highly unstable regions ($i.e.$ larger eccentricities) can still host systems with low $\Delta T_{\rm max}$~--~however, at a much reduced likelihood. The bottom panel of Figure \ref{fig:periodlog} emphasizes the mass dependence of $\langle \Delta T_{\rm max}\rangle$ even for stable scenarios. Stable systems with masses higher than $10^{19}$kg exhibit high perturbations which most probably will lead to collisions in the future. Alternatively, eccentricity does not show a prominent effect on stable, low-mass ($< 10^{19}$kg) systems.

Figure \ref{fig:maxaveminperiods} displays the temporal evolution of stable systems. We computed not only $\langle \Delta T_{\rm max}\rangle$ but also Max$(\Delta T_{\rm max})$ and Min$(\Delta T_{\rm max})$, being the maximum and minimum of $\Delta T_{\rm max}$ recorded in the 50 stable simulations. Max$(\Delta T_{\rm max})$, $\langle \Delta T_{\rm max}\rangle$ and Min$(\Delta T_{\rm max})$ are plotted at three different epochs throughout the simulations, six months, one year and two years after the start of the simulation. Figure \ref{fig:maxaveminperiods} reveals that there is not a noticeable variation of either Max$(\Delta T_{\rm max})$, $\langle \Delta T_{\rm max}\rangle$ or Min$(\Delta T_{\rm max})$ over time, meaning that stable systems do not evolve strongly with time and are robust regardless of sampling time. This finding is important because it ensures compatibility, and permits direct comparisons between simulations that last years and observational data that only spans months.

\begin{figure*}
\centering
	\includegraphics[width=2\columnwidth]{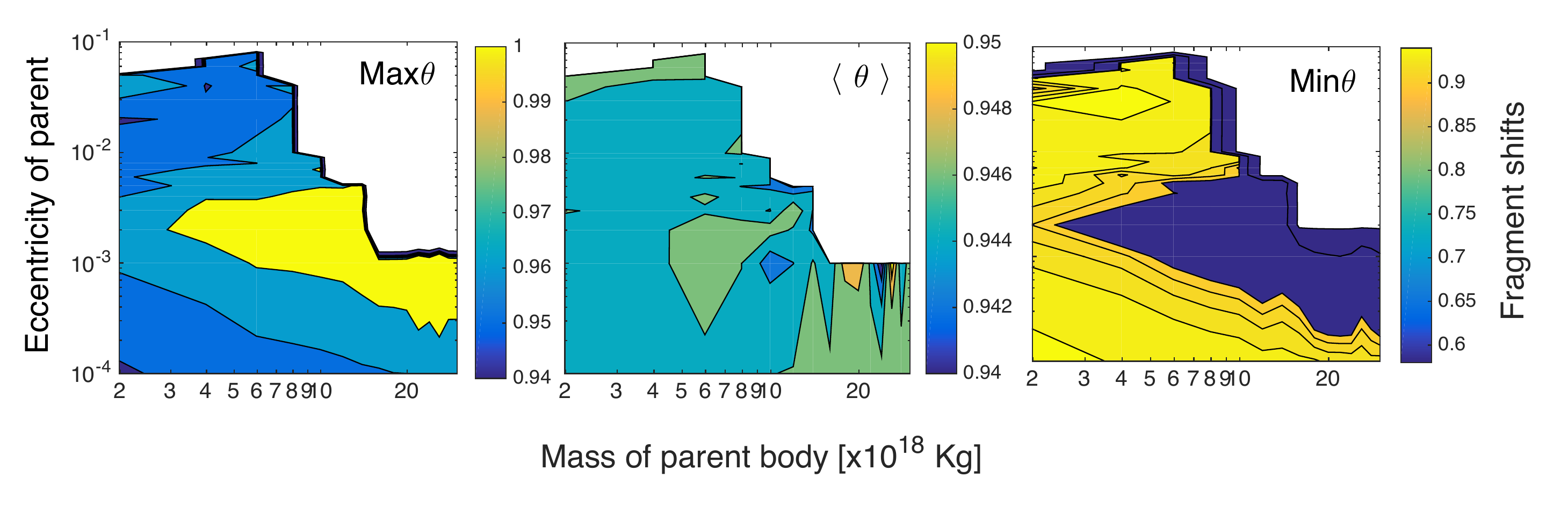}
    \caption{$\mathrm{Max}(\theta)$, $\langle \theta \rangle$ and $\mathrm{Min}(\theta)$ phase shifts in orbital cycles after 80 days (see Section \ref{sec:shifts}) for 50 stable simulations, with respect to mass and eccentricity of the parent body. $\mathrm{Max}(\theta)$ and $\mathrm{Min}(\theta)$ show the same structure as Figure \ref{fig:periodlog}.  $\langle \theta \rangle$ remains almost constant regardless of mass or eccentricity, meaning that on average, all fragments shift on phase in a similar way.}
    \label{fig:shifts}
\end{figure*}

\begin{figure*}
	\includegraphics[width=2\columnwidth]{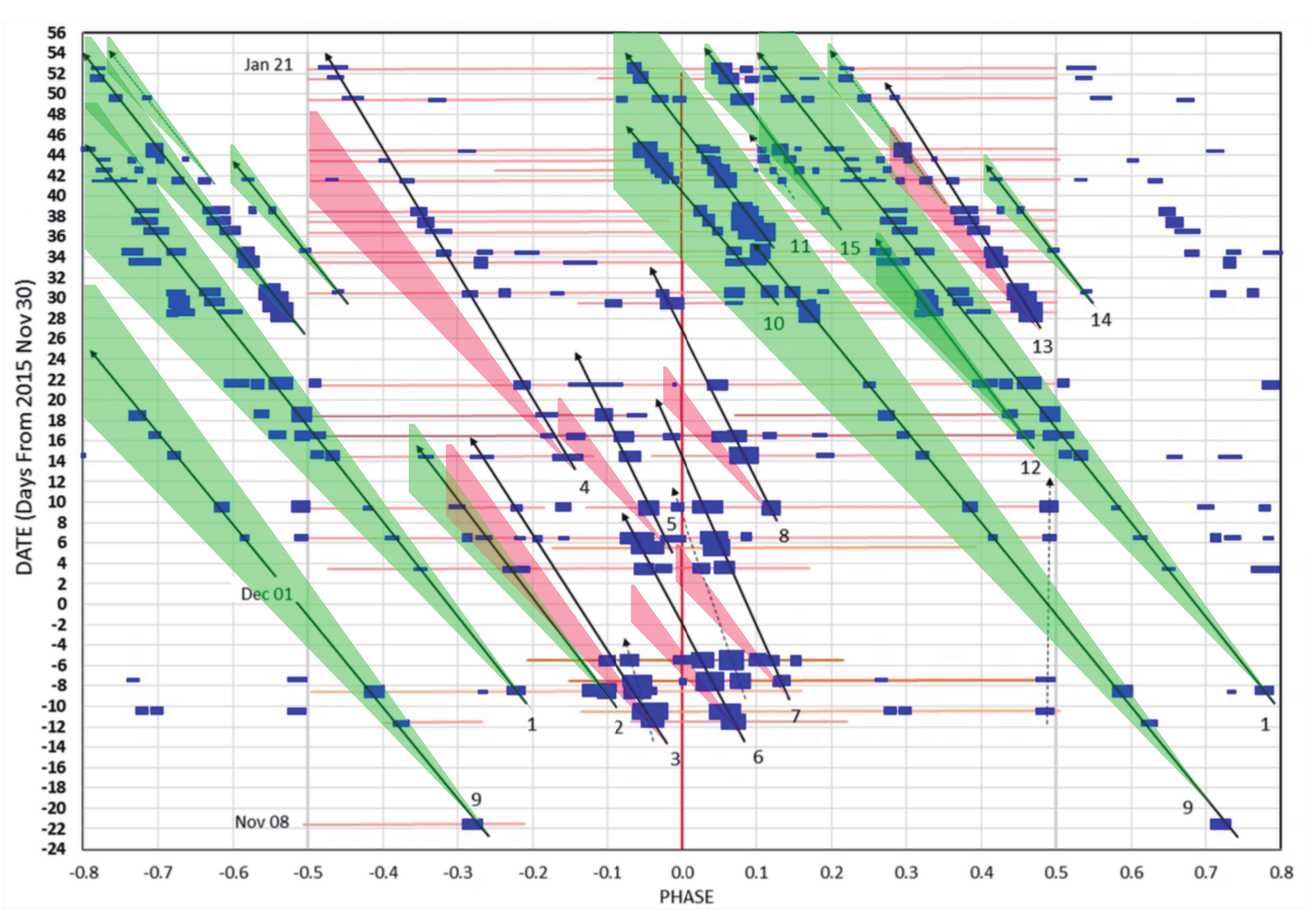}
    \caption{Superposition of Figure 6  of \citep{rappaport16} and the ranges of phase shifts predicted by our simulations, confined by $\langle \mathrm{Max}	(\theta) \rangle$ and $\langle \mathrm{Min}(\theta) \rangle$ (see Section \ref{sec:shifts}). Expected ranges are shaded green if the observational prediction \citep{rappaport16} fits with our simulations and red if it does not. Generally, computer predicted phase shifts suit observational data and every shaded range contains at least two linkable transits.}
    \label{fig:rappaportimage2}
\end{figure*}

\section{Phase Shifts} \label{sec:shifts}

The fact that the parent body and the fragments have different orbital periods
offers the possibility to determine phase shifts between them. By folding the transit signals of the fragments onto the parent's period, \cite{rappaport16} illustrated how the phase of the transit features shifts from night to night. Because $\Delta T_{\rm max}$ of stable systems remains almost constant in time, we used the stable simulations of Section \ref{sec:ecc} to calculate these phase shifts and obtain a direct comparison between observational data  and our simulations.

In order to determine an average phase shift, $\langle \theta \rangle$, we computed the ratio between the period of each fragment and the parent's one and defined the phase shift after 80 days as $\theta = (1-T_{\rm frag}/T_{\rm parent}) \times 24 {\rm h}/T_{\rm parent} \times 80 {\rm days}$. We double-checked our results by first averaging the period and then calculating phase shifts, which lead to very similar results, never differing more than 5\%.   

Figure \ref{fig:shifts} shows $\mathrm{Max}(\theta)$, $\langle \theta \rangle$ and $\mathrm{Min}(\theta)$ as a function of the eccentricity and mass of the parent body. We obviated phase shifts for systems exhibiting high $\langle \Delta T_{\rm max}\rangle$ (yellow area on Figure \ref{fig:instabilityHDD}) because observations suggest low period perturbations. Minimum and maximum phase shifts share a similar structure that resembles that of Figure \ref{fig:periodlog}. Mean phase shifts remain almost constant for all values of mass and eccentricity, meaning that on average, fragments tend to drift systematically regardless of their mass or eccentricity.

We averaged over all values of $\mathrm{Max}(\theta)$ and $\mathrm{Min}(\theta)$ shown in Figure \ref{fig:shifts} and obtained $\langle\mathrm{Max}(\theta) \rangle=0.9941$ and $\langle\mathrm{Min}(\theta) \rangle=0.8885$. We expect that most fragments experience phase shifts with respect to the parent that are contained between $\langle {\rm Max}(\theta) \rangle$ and $\langle {\rm Min}(\theta) \rangle$.

We superpose the range of phase shifts predicted by our simulations with the observations of \citet{rappaport16}. Fragments identified by \cite{rappaport16} that fall within the expected range are shaded in green, those that fall outside the predicted range are shaded in red. Overall, the agreement is remarkably good. We note that some of the transits interpreted by \cite{rappaport16} as a single fragment fall outside the predicted range of phase shifts (e.g. \#3--8), which illustrates both the difficulty in identifying individual fragments from the ever changing transits, and the possibility of additional dynamical processes among the fragments. 

\section{discussion} \label{sec:disc}

Constraining the mass of disintegrating objects orbiting a polluted white dwarf provides insights into the planetary system and the star itself. Asteroids have long since assumed to be source of the pollution \citep{graham1990,jura2003,beasok2013}, having been perturbed into the white dwarf disruption radius by external agents such as planets \citep{bonetal2011,debetal2012,frewen2014,ver2016b}, moons \citep{payne2016a,payne2016b}, comets \citep{alc1986,versha2014,sto2015} and/or wide binary companions \citep{bonver2015}. However, the size and mass distributions of the perturbed objects has remained unknown. 
To date, only lower limits on the parent body masses have been derived for a number of helium-atmosphere white dwarfs, based on the measured photospheric metal abundances and the model-dependent masses of the convection zones atmosphere white dwarfs (e.g. \citealt{koesteretal11, dufouretal12, giretal2012}. These mass estimates range from about $10^{16}$\,kg to $10^{23}$\,kg, which roughly encompasses the mass range of Saturn's moon Phobos to Jupiter's moon Europa.  However, these masses are averaged over the time scales on which the material diffuses out of the convection zone, typically $\simeq10^5-10^6$\,yr \citep{paquetteetal86, koester2009}, and hence may reflect the accretion of multiple, smaller parent bodies \citep{wyattetal14}.

WD\,1145+017 provides the first opportunity to directly determine the mass of the planetesimal undergoing disruption, and the maximum mass of the objects orbiting WD\,1145+017 derived here ($\sim 10^{20}$\,kg, see also \citealt{rappaport16}) is near the upper end of the range of metal masses in He-atmosphere white dwarfs (see Fig. 6 of \citealt*{veras2016rev}). For comparison \citet{xu16} derived the  mass of planetary debris accreted into the convection zone of WD\,1145+017 to be $\simeq6.6\times10^{20}$\,kg (over the last few diffusion time scales of a few $10^5$\,yr), i.e. the same order of magnitude as the amount of material still in orbit around the white dwarf. The dynamical mass estimate derived here, and by \citet{rappaport16} robustly corroborates the hypothesis of large rocky planetesimals surviving the red giant phase, and providing one possible source the metal pollution detected in many white dwarfs. 

For further comparison we used analytical estimates provided in equations 9-12 of \cite{vanderburg1}. Although crossing orbits do not always ensure instability, imposing such a condition provides a rough approximation of the eccentricity limit of the parent body of $\mathrm{eccentricity} < 10^{-3}$. Using the standard definition of the Hill Sphere it is also possible to obtain an estimate for the mass of the parent body. Assuming $\xi \sim 5-12$, a range which usually holds for distant, non-co-orbital objects \citep{chatetal2008,davetal2014,puwu2015,vergae2015,ver2016b}, we obtain a limiting mass of $M \simeq 5 \times 10^{19}$\,kg which is close to our numerically derived estimate of  $\sim 10^{20}$\,kg. The agreement with these analytical estimates adds further support to our numerical results.  

\section{Conclusions}  \label{sec:conc}

We have performed $N$-body simulations to derive constraints on mass and eccentricity of the planetary bodies orbiting WD\,1145+017. We found that either masses greater than $\simeq10^{20}$\,kg (0.1 the mass of Ceres) or orbits that are not nearly circular ($\mathrm{eccentricity}>10^{-3}$) increase the likelihood of dynamical instability over a timespan comparable to the baseline of the current set of observations. We also computed the phase shifts of the fragments with respect to the parent body, and found good agreement with the shifts measured by \cite{rappaport16}. 

Future work should include a detailed treatment of the actual disruption process, which would provide further insight into the physical properties of the disintegrating bodies, and allow an estimate of the expected duration of this phase.

\section*{Acknowledgments}

We thank the referee for all their valuable input.

The undergraduate exchange program of Engineering Physics from UPC (BarcelonaTech) made this research possible.

DV and BTG have received funding from the European Research Council under the European Union's Seventh Framework Programme (FP/2007-2013)/ERC Grant Agreement n. 320964 (WDTracer).

%%%%%%%%%%%%%%%%%%%%%%%%%%%%%%%%%%%%%%%%%%%%%%%%%%



\bibliographystyle{mnras}
\bibliography{library}

% Don't change these lines
\bsp	% typesetting comment
\label{lastpage}
\end{document}